# An Adaptive Honeypot Configuration, Deployment and Maintenance Strategy

Daniel Fraunholz, Marc Zimmermann, Hans D. Schotten

Intelligent Networks Research Group, German Research Centre for Artificial Intelligence,

Trippstadter Str. 122, 67663 Kaiserslautern, Germany

**Daniel.Fraunholz@dfki.de, Marc.Zimmermann@dfki.de, Hans_Dieter.Schotten@dfki.de**

*Abstract*— Since honeypots first appeared as an advanced network security concept they suffer from poor deployment and maintenance strategies. State-of-the-Art deployment is a manual process in which the honeypot needs to be configured and maintained by a network administrator. In this paper we present a method for a dynamic honeypot configuration, deployment and maintenance strategy based on machine learning techniques. Our method features an identification mechanism for machines and devices in a network. These entities are analysed and clustered. Based on the clusters, honeypots are intelligently deployed in the network. The proposed method needs no configuration and maintenance and is therefore a major advantage for the honeypot technology in modern network security.

## I. Introduction

Honeypots are an advanced concept in network security. The purpose of such a system is to gain information about intrusion attempts or intrusions of a resource. This information can be very different e.g. time, date, intruder IP address, intruder operating system or employed wordlists, exploits and commands after infiltration.

Honeypots are differentiated according to the following features: level of interaction, the deployment environment, resource type, services, and implementation. Since newer concepts also add context-aware features a state-of-the-art taxonomy can be defined as follows.

Level of interaction: The interaction level reaches from low-interaction honeypots, which emulate just the communication stack, to high-interaction honeypots, which run a real operating system.

Deployment environment: There are two general use cases for honeypots. The first one is the deployment in the wild, to analyse ongoing attack campaigns and upcoming zero-day-exploits. This concept is known as research honeypots and is widely used as sensor for antivirus database updates. The second deployment is behind the perimeter inside an enterprise or industrial network. In this deployment any interaction with the honeypot indicates a security breach. As they are closely to the production relevant entities, they are called production honeypots.

Resource type: Several concepts of deceptive resources are established. Server-side honeypots are systems with services in listening-mode which are waiting for incoming connections. Client-side honeypots are actively connecting to potential dangerous systems to investigate their intrusion attempts. Tokens are files or information which pretend to be accidentally leaked. For example, a password for an enterprise associated server. The usage of the password indicates a compromise of the source system.

Services: There is a set of services that is used by the honeypot system. To hinder the identification of honeypot systems the set of services must be chosen carefully, with respect to the expected services in the deployed environment.

Adaptability: Historically honeypots are static which means configuring, deploying and maintenance are manual tasks. Since their first days this is a major drawback of the technology [1]. Recent advances in machine learning and artificial intelligence revolutionize software adaptability. Honeypots that feature adaptability are called dynamic honeypots.

Implementation: Honeypots employing dedicated hardware are categorized as real honeypots. On the other hand, shared hardware honeypots are called virtual honeypots.

Regardless of the employed type, honeypots suffer from the trade-off between generalization and specification. On one hand the honeypot must attract the intruders interest in the honeypot resource and on the other hand it must be adapted to the current network status e.g. active machines and available services. In modern networks this task is hardly to achieve manually by administrators.

## II. Literature Review and Research Methods

There is a long history of honeypot research dating back to the first known application of a honeypot in 1990 [2].

Since this time a vast amount of honeypots was developed. But all of them need manual customization to avoid easy fingerprinting [3]. To overcome these issues context-aware features are employed. Context-awareness for honeypots comes in different shapes. One feature is the adaption to the intruder interaction [4–6]. But the proposed method is dedicated to the dynamic honeypot deployment research field which is in an early state compared to the original honeypot concept. There are several existing works in this field, which focus on different context-aware features for honeypots. Furthermore, the adaptive deployment includes different aspects like detailed learning of services [7], [8] and recognizing machines and their high-level features in a network [9].

Studies giving an overview were conducted in the field of context-aware honeypots [10], [11].

The proposed method is a contribution to the dynamic deployment of honeypots by identifying machines in a network and analysing their high-level features like the TCP-Stack-Fingerprint, services, uptime, MAC-Address and IP-Address.

### III. SOLUTIONS AND RESULT

#### A. Concept

The proposed system consists of four consecutive tasks, like depicted in Figure 1. The first task scans the network and generates a back-end database containing relevant information about the environment. In the second step machine learning methods are applied to analyse the back-end. As a result, the network is clustered in characteristic groups regarding the TCP-Stack-Fingerprint and open ports. Additionally, the distributions of IP-Addresses, MAC-Addresses and uptimes within a cluster are determined. The configuration data is generated by determining how many honeypots are generated per cluster. This depends on a cluster significance ranking system and the total number of honeypots. The honeypot generation is done by cloning the TCP-Stack-Fingerprint and open ports for the most significant machine within the cluster associated to the honeypot. The other features are set based on the deviation for every feature within the associated cluster. In the last step the configuration data is used to deploy the honeypots in the scanned network. As a maintenance feature the method can simply be reemployed after a time offset or another criterion.

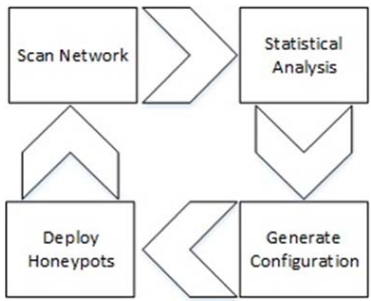

Figure 1. Schematic concept of the proposed method

#### B. Implementation

The implementation is based on Python. Several existing tools and libraries are used. Mentionable are Nmap as scanning engine, XML-Python-Lib as back-end and honeyd as deployment engine. As machine learning approach a modified version of k-means-clustering is used in combination with a binary Manhatten metric distance measurement between the feature vectors of two machines. The scheme is depicted in Figure 2.

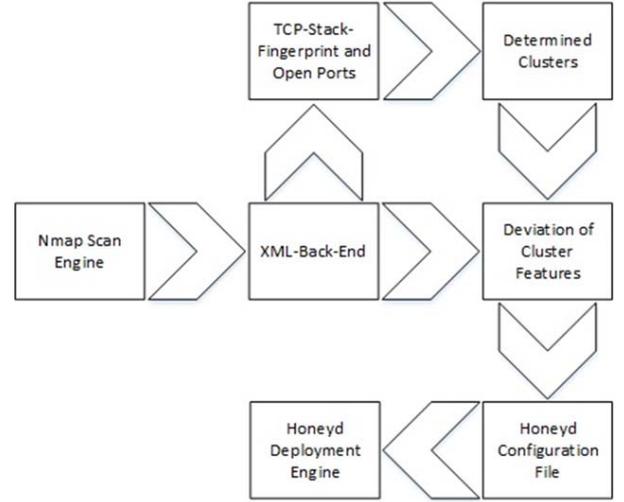

Figure 2. Scheme of implemented concept

*1) Scan engine:* As scanning engine nmap [12] is used. Nmap is a state-of-the-art tool for network scanning. It features a TCP-Stack-Fingerprint database and is able to determine all needed features for the algorithm.

*2) Data base format:* The format is defined as follows. The network consists of a number of entities. These entities have corresponding unique IDs and associated features that are used for network analyse purposes after the scanning phase.

*3) Clustering algorithm:* As clustering algorithm the k-means algorithm is employed. To overcome the drawbacks of k-means some adaptions to the algorithm are made. First the number of clusters is dynamic and determined in runtime. The heuristic is based on variance differences for several possible numbers of clusters. Second, to increase the probability to converge to the global minimum the clustering is done several times. As distance measurement the Manhatten metric is employed in a binary fashion. To determine the distance between two machines every TCP-Stack-Fingerprint feature and every open port is compared and the distance is increased by 1 for a mismatch in the feature value. The metric is defined as follows.

$$d_{(x,y)} = \sum_{n=0}^{f} |f_n^x - f_n^y| \qquad (1)$$

Where $d_{(x,y)}$ is the distance between two entities $x$ and $y$. $f_n$ is the feature vector of an entity and $n$ the feature number.

*4) Configuration and Deployment:* Honeyd [13] is able to deploy up to 65000 virtual honeypots in a network. Also several features like TCP-Stack-Fingerprint, IP-Address, MAC-Address and uptime can be defined. The configuration data is stored in a simple text file and can easily be generated in an automated process. Additionally, honeyd is able to emulate whole networks.

## C. Evaluation

The results of the algorithm strongly depend on the heuristics for the number of evaluations *H1* and the criteria of convergence for the number of clusters *H2*. This chapter depicts the process of determination of feasible parameters for both.

*1) H1 Number of Evaluations:* Since k-means-clustering is initiated randomly the variance within a cluster depends on the initialization and the number of evaluations. More evaluations give a higher possibility of finding the optimal clusters within the clustering process. But a higher number of evaluations increases the computation time. We analysed different evaluation numbers in respect of the total variance of all determined clusters and the number of determined clusters. The results are shown in Figure 3.

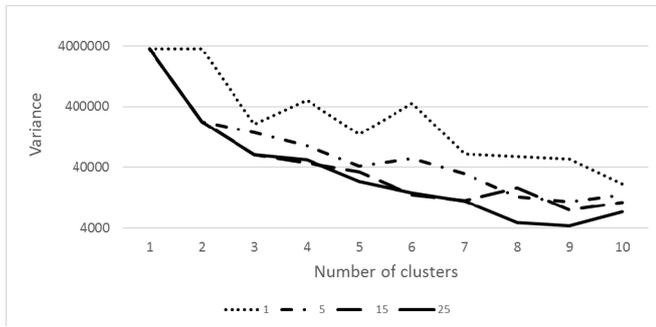

**Figure 3.** Total variance vs number of clusters for different numbers of evaluations

The expected result is that the total variance is monotonically decreasing. As it can be seen, for one evaluation the total variance strongly fluctuates. This is because the random initialization influences the final total variance. For a higher number of evaluations, the graph is closer to a monotonically decrease. Based on the results we chose an evaluation number of 15 for further experiments.

*2) H2 Number of Clusters*: For real world scenarios the number of cluster is not known before the algorithm starts. To determine a heuristic solution, we defined a data base with five known clusters and mutated the data base as follows.

$$Pr(\Omega \in [0,...,9]|f_{(n)}) = \beta, \text{für } f_{(n)} = \{F_{TCP}\} \quad (3)$$

$$Pr(\Omega \in [0,...,9]|f_{(n)}) = \begin{cases} \beta, \text{für } f_{(n)} \neq 0 \\ 0, \text{für } f_{(n)} = 0 \end{cases}, \text{für } f_{(n)} = \{F_{Ports}\} \quad (4)$$

Were $F_{TCP}$ describes the set of features regarding the TCP-Stack-Fingerprint and $F_{Ports}$ the feature set regarding the ports. $\beta$ is the mutation rate and gives the probability for a mutation for features. $f_{(n)}$ is the feature vector at feature number *n*. The feature value after mutation is $\Omega$.

After mutation we did 15 evaluations and calculated the total variance vs number of clusters graphs. As expected the graphs in Figure 4 can be considered as monotonically decreasing before convergence.

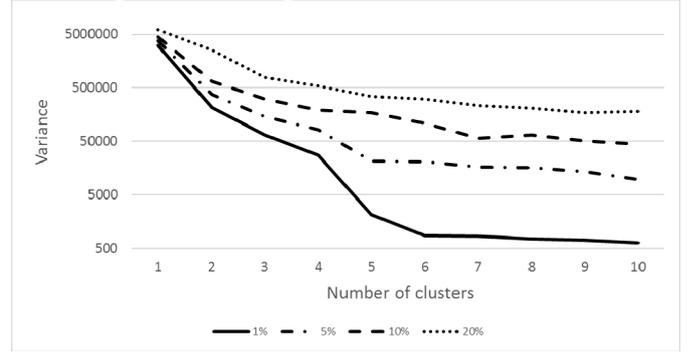

**Figure 4.** Total variance vs number of clusters for different mutation rates

Since the number of clusters is known in this scenario we analysed the results and determined a heuristic solution for the number of clusters convergence criteria. Different approaches like the k-means-elbow-criteria, polynomial and hyperbola criteria were tested. The best results were achieved for the following criteria.

$$\mu < \sigma_n - \sigma_{n-1} \quad (2)$$

Where $\mu$ is set to 0.68 and $\sigma_n$ is the total variance for $n$ clusters. The results are shown in Figure 5.

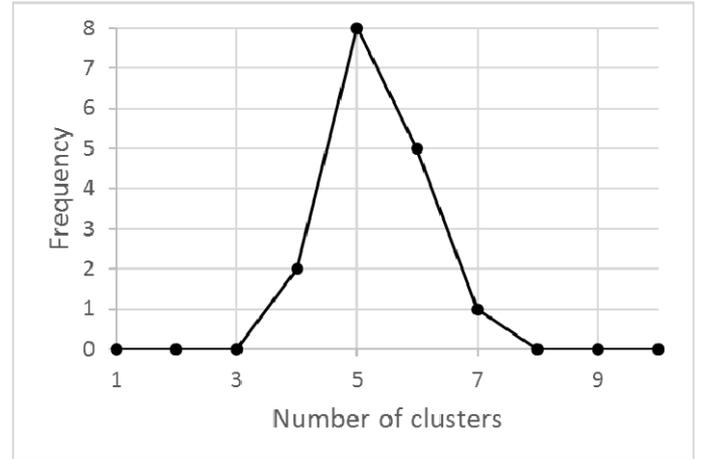

**Figure 5.** Occurrences for different number of clusters

## D. Real world scenario

To verify the number of evaluations an additional experiment was conducted. The convergence criteria for the number of clusters was tested for a real network. As expected the number of clusters converged for 15 evaluations as shown in Figure 6.

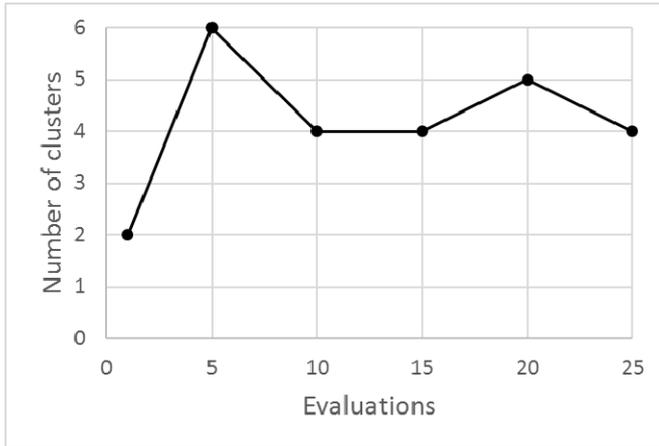

**Figure 6.** Determined number of clusters vs evaluations

After determining heuristic solutions and feasible parameters the algorithm is employed to cluster the network and determine the honeypot deployment.

### E. Results

Evaluations showed that the proposed algorithm is suitable to deploy virtual honeypots in a network. The deployed honeypots cannot be distinguished from real machines by their TCP-Stack, IP-Address, MAC-Address, uptime and open ports. Table 1 depicts the results for a real world scenario in a development environment. As shown the Raspberry Pi systems which all run Raspbian are correct clustered in cluster 1. The Windows 10 machines are all in cluster 2. Our experiments show that our distance metric often determines TP-Link switches and Windows 10 in one cluster. The other Unix system in the Windows 10 cluster is a FreeBSD based PFsense distribution running as firewall.

Table 1  RESULTS OF THE CLUSTERING ALGORITHM FOR A REAL WORLD CLASS C SUBNET

| System | Count | Cluster 1 | 2 | 3 | 4 | 5 |
|---|---|---|---|---|---|---|
| PC Ubuntu | 4 | 1 | 0 | 2 | 1 | 0 |
| TP-Link Switch | 2 | 0 | 1 | 0 | 1 | 0 |
| PC Windows 10 | 7 | 0 | 7 | 0 | 0 | 0 |
| Cisco Switch | 2 | 0 | 0 | 0 | 1 | 1 |
| Android | 1 | 0 | 0 | 0 | 1 | 0 |
| Raspberry Pi | 11 | 11 | 0 | 0 | 0 | 0 |
| Other Unix | 4 | 0 | 1 | 2 | 0 | 1 |

Based on this results the honeypot will be configured by the following definitions.

*1) Choice of cluster:* For the first honeypot deployment the cluster with the highest number of entities is chosen. For further deployments the number of entities in the biggest cluster is divided by 2 for every deployed honeypot in this cluster. This ensures a broad deployment in all clusters but also considers the influence of different cluster sizes.

*2) TCP-Stack:* Since in the presented implementation the centroid for any cluster corresponds to a real system, the TCP-Stack of the honeypot is equal to the TCP-Stack of the centroid system of the determined cluster.

*3) IP Address:* Within the corresponding cluster the distribution of IP addresses is estimated and the honeypot IP address is calculated in respect to the estimated distribution.

*4) MAC Address:* The first three bytes are equal to the most prevalent vendor in the cluster and the last three are determined randomly.

*5) Uptime*: The uptime is determined by calculating the mean value for all uptimes within the cluster. Occurring delays between scan time and deployment time are corrected.

### IV. CONCLUSIONS

This paper proposed an algorithm for context-aware honeypot deployment in unknown or dynamic networks. This feature mitigates the most prevalent barrier to a large-scale honeypot distribution. We showed that the algorithm is currently operational and able to increase threat intelligence with a minimum of configuration and maintenance. Further research will investigate more high level features like service behaviour for the deployment. By learning and imitating services in the network the intruder is unable to distinguish honeypots and real machines even with high interaction.

Additionally, in this paper an enhanced taxonomy for honeypots was presented. This taxonomy considers the adaptability features of a system, which is not present in other state-of-the-art taxonomies.

Future research will also consider the generalization problem. The basic idea is to make the honeypot more attractive to intruders. This can be achieved by lowering the honeypot security level and increasing the expected value for the intruder.

### ACKNOWLEDGMENT

This work has been supported by the Federal Ministry of Education and Research of the Federal Republic of Germany (Foerderkennzeichen KIS4ITS0001, IUNO). The authors alone are responsible for the content of the paper.

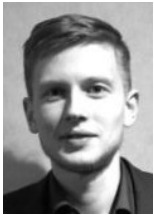
**Daniel Fraunholz** is a Researcher and Ph.D. candidate at the Intelligent Networks Research Group at the German Research Center for Artificial Intelligence since 2015. Born in Stuttgart, Germany in 1992, he received his Bachelor and Master degree from Heilbronn University of Applied Sciences in Electronic Systems Engineering in 2014, respectively 2015. His major research interests are network security, intrusion detection and honeypots.

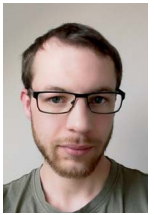
**Marc Zimmermann** became a Junior Researcher at German Research Center for Artificial Intelligence in the Intelligent Networks Research Group in 2016. He was born in Zweibrücken, Germany, in 1992. Zimmermann got the Bachelor of Science degree in Media and Communication Techniques from Technical University of Kaiserslautern, Germany, in 2016. Currently he is a candidate for the Master of Science degree in Media and Communication Technology. His resent research interests are in network and system security and analysis of intruder behavior via honeypots.

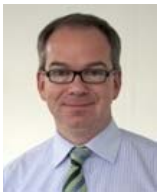
**Hans Dieter Schotten** is Scientific Director of the Intelligent Networks Research Group at the German Research Center for Artificial Intelligence. Since 2007 is also Professor at the Technical University of Kaiserslautern. He received his Ph.D. from the RWTH Aachen University in Electrical Engineering. He has worked as Senior Researcher at Ericsson and as Director for Technical Standards at Qualcomm.